\documentclass[epj]{svjour}
\usepackage{graphics}
\usepackage{etoolbox}
\usepackage{soul}

\newcommand*{\myfont}{\fontfamily{lmss}\selectfont}
\AtBeginEnvironment{enumerate}{\myfont}
\begin{document}


%
\title{Collective behavior of Vicsek particles without and with obstacles}
\author{Raul Martinez\inst{1,2,3} \and Francisco Alarcon\inst{1,3} \and Diego Rogel Rodriguez \inst{1,3} 
\and Juan Luis Aragones\inst{2} \and Chantal Valeriani\inst{1,3}
}
\titlerunning{Collective behavior of Vicsek particles without and with obstacles}
\authorrunning{Martinez \textit{et al.}}    
\offprints{cvaleriani@ucm.es}          
\institute{Departamento de estructura de la materia, f\'{\i}sica t\'{e}rmica y electr\'{o}nica, Facultad de Ciencias  F\'{\i}sicas, Universidad Complutense de Madrid,
28040 Madrid, Spain \and IFIMAC, Facultad de Ciencias, Universidad Aut\'{o}noma de Madrid, Ciudad Universitaria de Cantoblanco, 28049, Madrid, Spain \and
GISC - Grupo Interdisciplinar de Sistemas Complejos, Madrid, Spain
}
%
%
\abstract{
In our work we have studied a two-dimensional suspension of  finite-size Vicsek hard-disks, whose time evolution 
follows an event-driven dynamics between subsequent time steps. 
Having compared its collective behaviour with the one expected for a system of 
scalar Vicsek point-like particles, 
we have analysed the effect of considering  two possible bouncing rules between the disks: a Vicsek-like 
rule and a pseudo-elastic one, focusing on the  order-disorder transition.
Next, we have added to the two-dimensional suspension of  hard-disk Vicsek  particles
 disk-like passive obstacles of two types: either fixed in space or moving according to the same event-driven dynamics. 
 We have performed a detailed analysis of the  particles' collective behaviour observed for both fixed and moving obstacles. In the fixed obstacles case, we have observed formation of clusters at low noise, in agreement with previous studies. When using moving passive obstacles, we found that that order of active particles is better destroyed as the drag of obstacles increases. In the no drag limit an interesting result was found: introduction of low drag passive particles can lead in some cases to a more ordered state of active flocking particles than what they show in bulk. 
%
} 
\maketitle
\section{Introduction}

Active matter deals with out-of-equilibrium systems of self-driven units (active particles) each able to convert energy (either stored or from its surroundings) into directed motion\cite{schweitzer2003brownian,ba,marchettirev,menonbook,malloryrev,ramaswarev,catestailleur}. Due to the particle-particle interaction and to the particles' interplay with the surrounding medium, active particles are capable of  a characteristic collective behaviour, 
which has raised great interest over the last few decades, both theoretically and experimentally \cite{th,mh,vc,ap,deseigne2010collective,weber2013long}. 

One of the simplest numerical models used to understand order-disorder out-of-equilibrium transitions
 and  features of the collective motion   in active suspensions \cite{gregoire2004onset,chate2008collective,ginelli2016physics}
 is the Vicsek model \cite{vicsek1995novel}.
Despite the simple interaction rules acting between self-propelled point-like particles,
 the  model has proven capable of capturing the
 behavior of  systems with emergent collective motion,
ranging from flocks of birds to microorganisms cluster formation.  

In order to describe more realistic systems, 
such as the formation of schools of fish \cite{gautrais2012deciphering} or to model bacterial motion  \cite{gregoire2001active,chate2008modeling,chantalSM2011}, 
modified versions of the Vicsek model have been proposed,  where self-propelled hard-disks take the place of self-propelled point-like particles.  

Several studies have focused on the collective behaviour of self-propelled particles in crowded environments \cite{ba}, where chemotactic particles show anomalous dynamics on a percolating cluster \cite{schilling2017clearing}, while clustering has been observed when chemotactic particles move in a porous medium \cite{marsden2014chemotactic} or in bulk \cite{pohl2014dynamic}.
   At the same time, several studies have been reported with  
   self-propelled particles  interacting with  obstacles   
 either   fixed  in space \cite{chepizhko2013diffusion,chepizhko2015active}
or moving due to the collisions with  the active particles  \cite{chantalSM2011,yllanes2017many}.
 

Inspired by the above mentioned works, we propose to study a two dimensional dilute suspension of self-propelled Vicsek 
hard disks and compare them to their point analog,   in the presence of either fixed or mobile hard-disks obstacles. 
To start with, we describe the event-driven algorithm to simulate hard-disks Vicsek-like particles, considering two possible 
bouncing rules between particles: a Vicsek-like  and a pseudo-elastic one. To establish the effect of adding a finite area to 
the self-propelled disks, we will  compare the results with the the Vicsek-points model analog, focusing on the order-disorder transition. 

Next ,we will simulate Vicsek particles (either point-like or hard disks)  interacting with fixed disk-like obstacles, 
considering the case where  particles' size is much smaller of that of the obstacles. 
In either case our results agree with the ones presented by Chepizhko and Peruani \cite{chepizhko2015active}.
To conclude, we will let the obstacles move, and study the resulting phenomena.

The manuscript is organised as follows. In section \ref{Sec:Simulation_Details} we present the numerical details 
including  bulk systems and systems with obstacles. 
In section \ref{Sec:Results}, we present the obtained results and 
draw our conclusions in section \ref{Sec:Conclusions}. 

\section{Model and simulation details} \label{Sec:Simulation_Details}

To start with, in \ref{s1} we  introduce the original version of the Vicsek model  \cite{vicsek1995novel,gregoire2004onset}.
Next, in section \ref{s2} we propose the Vicsek model for hard-disks, thus presenting two bouncing rules responsible for the interactions between them: 
a Vicsek-like rule, which aligns the velocities of the colliding particles, inspired by the Vicsek model update rule; and 
a pseudo-elastic rule, inspired by the possibility of the existence of an inertial dynamic at short times.  

Having discussed the bulk system, we present the numerical details for the system with fixed obstacles. 
In section \ref{s3}, the interaction between particles and fixed obstacles is discussed.
 Two bouncing rules are considered: one that aligns particles with the wall and another that performs an elastic (or mirror) bouncing. 
  The elastic bouncing rule  is inspired by the possibility of the existence of an inertial dynamic at short times,
   while the aligning  bouncing rule is inspired by hydrodynamic interactions with walls.

In section \ref{s4}, we present the numerical details for the system with moving obstacles: 
in our model, passive particles are desplaced by the collisions 
  with other particles and   slowed down by a drag force  (to model the interaction with the fluid).
  Two bouncing rules are considered:
  one describing collisions between  two obstacles and another one for the obstacle - particle collision. 
Given that for obstacles the constant speed condition does not have to be imposed, 
we consider 1) an elastic bouncing rule for the obstacle-obstacle bouncing rule, 
and 2)  a pseudo-elastic bouncing rule for the active particle-obstacle.s 


To conclude the method section, we will present the chosen parameter range \ref{s5}.

\subsection{Self-propelled Vicsek point-like particles}\label{s1}

In the scalar Vicsek model \cite{chate2008collective,gregoire2004onset}, the  system consists of $N$ identical point particles 
in a two dimensional  square box of edge $L$. 
All particles propel at a constant speed $v$, thus the velocity $\vec{v}_i$ of the $i$th-particle at time $t$ 
is completely determined by its direction of motion $\theta_i$
\begin{equation}
\vec{v}_i=v (\mathbf{\hat{\i}} \cos \theta_i(t) + \mathbf{\hat{\j}} \sin \theta_i(t)).
\end{equation}
 Every  time step $\Delta t$, $\theta_i$ is updated  as  follows:

\begin{enumerate}
\item We identify all neighbouring particles $j$ of the $i$th-particle as those at a  distance smaller than $R$:
$|\vec{r}_i(t)-\vec{r}_j(t)|<R$, where $R$ is the interaction range (metric Vicsek model).
\item We compute  $\theta^m_i(t)$  as the mean direction of all  neighbouring particles of the $i$th-particle.
\item The new direction of the $i$th-particle is $\theta_i(t+\Delta t)=\theta^m_i(t)+\eta\zeta_i(t)$, where 
$\zeta_i(t)$ is a white noise  ($\langle \zeta(t) \rangle=0$, $\langle \zeta_i(t) \zeta_j(t') \rangle = \delta_{ij} \delta(t-t')$, and $\zeta_i \in [-\pi,\pi]$) 
and $\eta$ the noise strength  ($\eta \in [0,1]$, set at the beginning of each  run). 
\item Each particle's position  is then updated according to $\vec{r}(t+\Delta t)=\vec{r}(t)+\vec{v}(t+\Delta t)\Delta t$.
\end{enumerate}

All simulations are performed with  periodic boundary conditions (unless  stated otherwise).  
We make use of reduced units, with $\Delta t$ as the time unit and $R$ the  length unit
 ($\Delta t = 1$ and  $R=1$). In all  simulations, $v\Delta t < R$.
We initialise the system with random positions and velocities, $\vec{r}$ and $\theta$ for each particle, and let it evolve 
 until  steady state is reached. All results presented  (including snapshots) are taken from the steady state.
 We fix the particles' density $\rho = N/L^2$ to be  $\rho=1.953$ throughout our simulations.


To measure the total alignment within the system, we compute the 
 order parameter $\nu$ \cite{gregoire2004onset}:
\begin{equation}
\label{ordpar}
\nu = \frac{1}{N}\sqrt{\left(\sum_{k=1}^N\sin (\theta_k)\right)^2+\left(\sum_{k=1}^N\cos (\theta_k)\right)^2} 
\end{equation}
\noindent
Whenever $\nu=0$  particles  are  disordered and  move randomly in every direction; whereas when $\nu=1$ particles  are  
totally aligned and  move as a flock. 

To study particles aggregation, we detect all clusters in the system via  the following criterion: 
at a given time step in the steady state, two particles belong to the same cluster if their distance  is smaller than the interaction range $R$. 
The largest cluster $n_c$ will just be the cluster with the largest number of particles. 

\subsection{Self-propelled  Vicsek hard-disks}\label{s2}

\begin{figure}
\centering
\resizebox{0.4\textwidth}{!}{%
\includegraphics{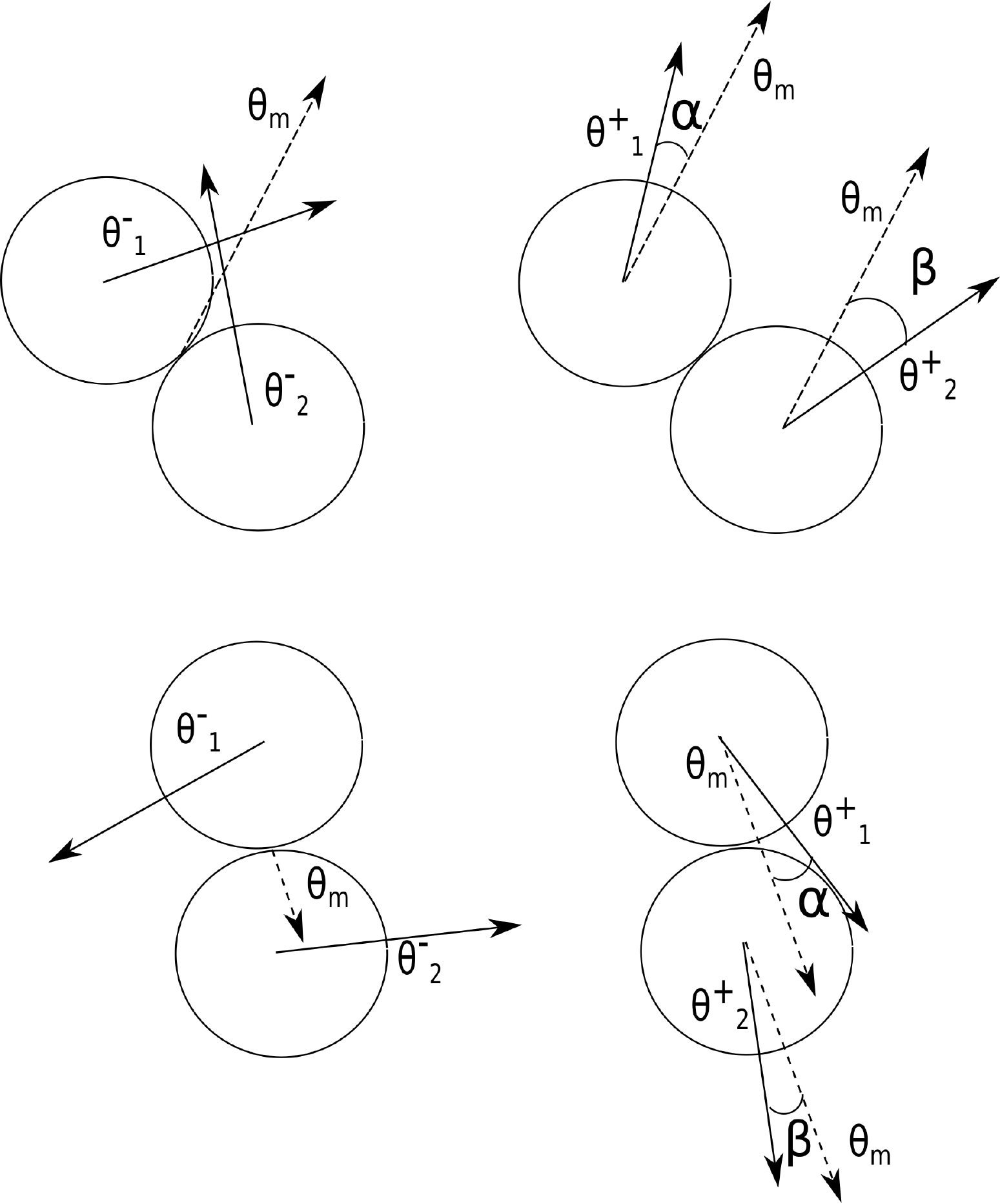}
}
\caption{Top row: before (left) and after (right) the collision according to the Vicsek-like bouncing rule. Bottom row: another  case  before (left) and after (right) the collision according to the Vicsek-like  rule. The Vicsek-like bouncing rule consists in: 1) Calculating the mean direction between both particles ($\theta_m$) before the collision, and 2) Adding a noise to it (angles $\alpha$ and $\beta$) after the collision. 
}
\label{chdos}       
\end{figure}

\begin{figure}[h!]
\centering
\resizebox{0.4\textwidth}{!}{%
\includegraphics{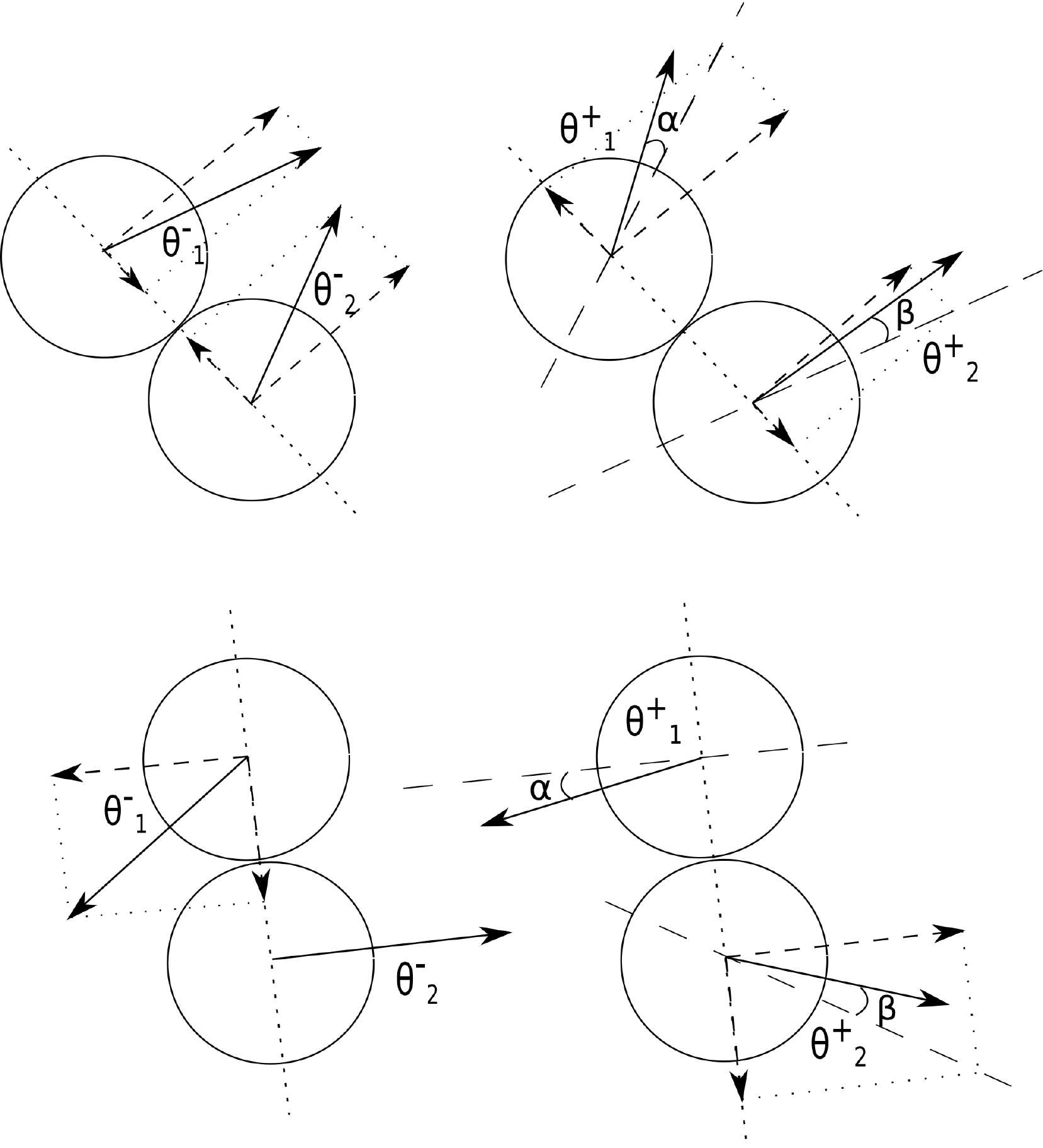}
}
\caption{Top row: before (left) and after (right) the collision according to the pseudo elastic bouncing rule. Bottom row: another  case  before (left) and after (right) the collision according to the pseudo elastic  rule. The pseudo elastic bouncing rule consists in: 1) Applying an elastic bouncing rule between two hard disks, 2) Rescaling the magnitude of the resulting velocities to match the constant speed condition, and 3) Changing the direction of the resulting velocity due to a noise (angles $\alpha$ and $\beta$) after the collision. }
\label{chpsel}       
\end{figure}


 When considering finite-size Vicsek  disks (diameter $\sigma$)
 at  packing fraction  $\phi = (N\pi\sigma^2)/(4L^2)$.

 Disks' velocities are updated at intervals of $\Delta t$ as prescribed by  the Vicsek model. 
 However, between these intervals they follow an event-driven algorithm  \cite{alder1959studies} 
 \footnote{Care must be taken when simulating 
at low $\eta$, where one should expect the formation of dense regions  (given that  the point-like Vicsek model tends to form clusters). 
This might cause numerical errors in the event-driven even algorithm even at low packing fractions. 
To solve this numerical issue the error-correction mechanism exposed in \cite{bannerman2014stable} has been successfully implemented.}
  and can interact with each others via  two different bouncing rules: a Vicsek-like  and a pseudo-elastic-like bouncing rule.


\textbf{The Vicsek-like bouncing rule} tries to mimic a Vicsek  interaction between two disks.
As schematically represented in  figure \ref{chdos}, it is applied as follows:

\begin{enumerate}
\item Given the velocities just before the collision of two particles, say $\vec{v}^-_1$ and $\vec{v}^-_2$
 for particle $1$ and $2$ respectively, $\theta^m$ is computed as $\theta^m = 
(\theta^-_1 +  \theta^-_2) /2$ (left panels in figure \ref{chdos}, top and bottom).
\item Let $\vec{r}_{rel}$ be $\vec{r}_2-\vec{r}_1$ and $\theta_{rel}$ 
the direction of $\vec{r}_{rel}$. 
\begin{enumerate}
\item If $\sin(\theta^m - \theta_{rel})>0$ then $\theta ^+_1=\theta^m + \eta\zeta/2$
and $\theta ^+_2=\theta^m - \eta\zeta/2$.
\item if  $\sin(\theta^m - \theta_{rel})<0$, 
$\theta ^+_1=\theta^m - \eta\zeta/2$ and $\theta ^+_2=\theta^m + \eta\zeta/2$  .
\end{enumerate}
\end{enumerate}

$\zeta\in [0,\pi]$ 
is a white noise and $\eta$ is the Vicsek model's noise strength.
All noise  terms correspond to $\alpha$ and $\beta$ in the right panels in figure \ref{chdos} (top and bottom).
Noise is introduced to better reproduce a Vicsek-like interaction 
and to guarantee that particles move away from each other after bouncing. \\




\textbf{The pseudo-elastic bouncing rule} tries to mimic an elastic collision between particles. Given that the  Vicsek self-propelled disks have a constant speed, 
we use a modified elastic bouncing rule   
as schematically represented in  figure \ref{chpsel}. 

\begin{enumerate}
\item We consider a frame consisting of the disk 1-to-disk 2  center-to-center direction (transversal, T) and its perpendicular direction 
(longitudinal, L), and decompose the particles' velocities into 
$\vec{v}^-_1 = (v^-_{1L},v^-_{1T})$, $\vec{v}^-_2 = (v^-_{2L},v^-_{2T})$. 
\item As in an elastic bouncing, after the collision we exchange particles' transversal components 
$\vec{v}_1^{\hspace{0.1cm}e}=(v_{1L}^-,v_{2T}^-)$, $\vec{v}_2^{\hspace{0.1cm}e}=(v_{2L}^-,v_{1T}^-)$.
\item Having computed $\theta_i^e$  ($i=1,2$) (direction of motion of $\vec{v}_i^{\hspace{0.1cm}e}$) 
 the next direction is $\theta_i^{+}=\theta_i^e+\eta\zeta$, where $\zeta\in [-\pi,\pi]$ 
 is a white noise.
\end{enumerate}

\subsection{Fixed disk-like obstacles}\label{s3}

To model a system consisting of self-propelled particles with fixed obstacles, we introduce $N_{d}$   finite-diameter 
$\sigma_d$ disks  placed 
at random in the simulation box of edge $L$, with a minimum center to center distance
 of  $(3/2)\sigma_{d}$ among them. Packing fraction of the obstacles is defined as $\phi_d = (N_d\pi\sigma_d^2)/(4L^2)$.

Disks' velocities are updated at intervals of $\Delta t$ as prescribed by  the Vicsek model 
 and follow an event-driven algorithm  \cite{alder1959studies} between these  time intervals.
  \footnote{Care must be taken given that an event driven algorithm used to simulate a suspension of hard disks could enter an "infinite" loop 
 even at low packing fractions. As s possible numerical way to avoid it, we suggest 
 to add an arbitrary  anti-loop rule, consisting of  setting to 0  the velocity of a particle experiencing more than 1000 
 collisions and keeping it set to 0 until the next iteration (even when other particles collide with it).
 We should mention that we have checked that the chosen number of iterations (1000) does not affect the dynamics of our system 
(allowing to distinguish between a particle in an infinite loop and  a particle experiencing numerous collisions).
 }

\begin{figure}[h!]
\centering
\resizebox{0.25\textwidth}{!}{%
\includegraphics{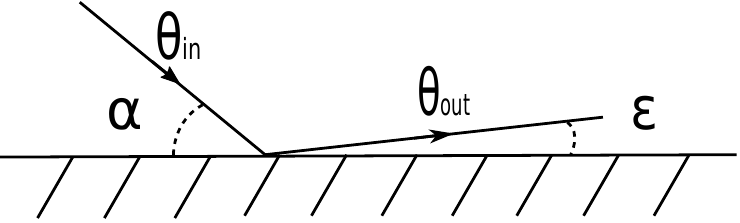}
}
\resizebox{0.25\textwidth}{!}{%
\includegraphics{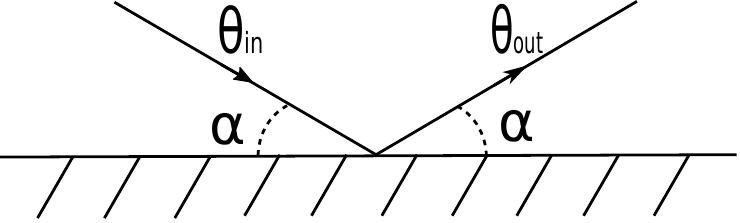}
}
\caption{Top:  aligning with the wall bouncing rule.  Bottom: elastic bouncing rule. Both without adding any noise. 
The straight line of the obstacle represents  the tangent to the obstacle in the point where the collision takes place.
}
\label{bouncing}     
\end{figure}

When a Vicsek particle collides with an obstacle, the point of intersection is calculated, and then a
 bouncing rule is applied.  
We consider two possible bouncing rules: a  particles-{\bf aligning} with the walls (top sketch in figure \ref{bouncing})
and an {\bf  elastic-bouncing} rule (bottom sketch in figure \ref{bouncing}).

The {\bf aligning} with the walls bouncing  rule consists in the following steps:
\begin{enumerate}
\item The angle $\theta_{in}$ formed by the incoming particle and the tangent of the obstacle is computed.  
\item $\theta_{out}$ (direction of motion  after the collision relative to the tangent of the obstacle) 
is set to 
\begin{enumerate}
\item $\theta_{out}=\epsilon$\hspace{1.45cm} if\hspace{0.2cm} $\cos(\theta_{in})>0$ 
\item  $\theta_{out}=\pi-\epsilon$ \hspace{0.8cm}if\hspace{0.2cm} $\cos(\theta_{in})<0$
\end{enumerate}
being $\epsilon\in [0,\eta\pi]$   
 a delta correlated white noise and $\eta$ the Vicsek model's noise constant.
\end{enumerate}

Whereas  the {\bf elastic-bouncing} rule consists in the following steps:

\begin{enumerate}
\item The angle $\theta_{in}$   formed by the incoming particle and the tangent with the obstacle is computed.
\item $\theta_{out}=-\theta_{in}+\epsilon$ (where $\theta_{out}$  is the angle formed  with the tangent to the obstacle) 

 $\epsilon$ is a delta correlated white noise with $\epsilon\in [-\eta\pi,\eta\pi]$ 
and $\eta$ the Vicsek model's noise constant.
\end{enumerate}

\subsection{Moving disk-like obstacles}\label{s4}


To model a system consisting of self-propelled particles with moving obstacles, we introduce $N_{d}$   finite-diameter 
$\sigma_d$ disks  (larger than the active particles) placed at random 
(being $M=\sigma_d^2 /\sigma^2$). We let  evolve the passive disks with time according to 
an over-damped equation of motion, where 
passive particles are slowed down by scaling their speed at regular time steps $\Delta t'$
\begin{equation}
 \vec{v}(t + \Delta t') = (1-\gamma) \vec{v}(t)
\end{equation}
being $\gamma$ the drag coefficient.

When two passive obstacles collide, we apply a purely elastic bouncing rule in which the linear momentum is conserved
(not having to maintain their speed after collision, unlike Vicsek particles). In what follows, we will 
 consider that the mass of the passive particles is proportional to their occupied area.

 The {\bf obstacle-obstacle} collision rule is the following:
 \begin{enumerate}
 \item We consider a frame consisting of the disk 1-to-disk 2  center-to-center direction (transversal, T) and its perpendicular direction 
(longitudinal, L), and decompose the particles' velocities into 
$\vec{v}^-_1 = (v^-_{1L},v^-_{1T})$, $\vec{v}^-_2 = (v^-_{2L},v^-_{2T})$. 
\item As in an elastic bouncing, after the collision we exchange particles' transversal components 
$\vec{v}_1^{\hspace{0.1cm}e}=(v_{1L}^-,v_{2T}^-)$, $\vec{v}_2^{\hspace{0.1cm}e}=(v_{2L}^-,v_{1T}^-)$.
\end{enumerate}

The only difference between this bouncing rule and  the pseudo-elastic one is that veocity is not rescaled 
(given that  passive particles do not have the constant speed constraint).

 The {\bf active particles-obstacle} collision rule, designed to maintain the constant speed of  the active particle,  is the following:
\begin{enumerate}
\item We consider a frame consisting of the active (A)-to-passive (P) center-to-center direction (transversal, T) and its perpendicular direction 
(longitudinal, L), and decompose the particles' velocities into 
$\vec{v}^-_A = (v^-_{AL},v^-_{AT})$, $\vec{v}^-_P = (v^-_{PL},v^-_{PT})$.
\item After collision, transversal components are modified as:
$v^e_ {AT}=(v^-_{AT}(1-M)+2Mv^-_{PT})/(M+1)$ \\
$v^e_ {AP}= (v^-_{PT}(M-1)+2v^-_{AT})/(M+1)$\\
so that \\
$\vec{v}_A^{\hspace{0.1cm}e}=(v_{AL}^-,v_{AT}^e)$ \\
 $\vec{v}_P^{\hspace{0.1cm}e}=(v_{PL}^-,v_{PT}^e)$ (elastic bouncing). 
\item $\vec{v}_P^{+}=\vec{v}_P^{\hspace{0.1cm}e}$ is the new velocity of the passive particle.
\item Having computed the direction of motion of $\vec{v}_A^{\hspace{0.1cm}e}$, $\theta_A^e$,  
 the new direction is \\
 $\theta_A^{+}=\theta_i^e+\eta\zeta$\\
 where $\zeta\in [-\pi,\pi]$ 
 is a white noise.
\end{enumerate}


\subsection{Parameter range}\label{s5}

When dealing with point-like Vicsek particles, we have simulated  only one number density of active particles  ($\rho=1.953$),  two system sizes ($L=32$ and $L=64$) and propulsion speeds within $v\in [0.1,0.5]$. When including  disk-like obstacles, we have simulated  a packing fraction of $\phi _d =0.0707$  
with a  number density of active particles ranging from $\rho_d=0.015$ to $\rho_d=0.25$. 
 
 When considering disk-like self-propelled  particles, we have considered a wide range of packing fractions. 
 When including  disk-like obstacles,  we restricted ourselves to very low packing fractions: $\phi < 0.06$, given that 
a constant speed  combined with an event-driven dynamics 
makes numerical simulation at higher $\phi$ very inefficient, 
as particles will more likely cluster  in the presence of  obstacles, while continuously colliding with their neighbours. 








\section{Results} \label{Sec:Results}

The results are presented as follows. We will start with presenting 
the order-disorder transition in the Vicsek disk-like particles as compared to the point-like one (\ref{res1})
Next, we consider the effect of inserting  disk-like obstacles on the collective behaviour of point-like Vicsek particles, when the bouncing rule 
is either aligning (\ref{res2}) or elastic (\ref{res3}) 
To conclude, we will study  the effect  on the collective behaviour of  disk-like Vicsek particles of fixed obstacles
 (\ref{res4}), comparing  several bouncing rules, and of mobile obstacles (\ref{res5}).

\subsection{The bulk system: comparison between Vicsek point-like  and disk-like particles}\label{res1}

The collective behaviour of a two-dimensional suspension of point-like Vicsek particles has been  widely studied in the literature\cite{chate2008collective,ginelli2016physics,gregoire2004onset}. However, as far as we are aware, 
not too much effort has been out into understanding 
 the effect that considering excluded volume interactions between finite size Vicsek particles might have on the order-disorder transition. 

In our work, we quantify the differences in 
 the order-disorder transition of point-like and  finite-size disk-like Vicsek particles. 
 When comparing the two systems, we  use the same values of $N$, $L$ and $v$,  
while modifying the value of the hard-disks diameter $\sigma$   (and the packing fraction, being  
 $\phi\to 0$ the limit of the point-like Vicsek model).

\begin{figure}
\centering


\resizebox{0.4\textwidth}{!}{%
\includegraphics{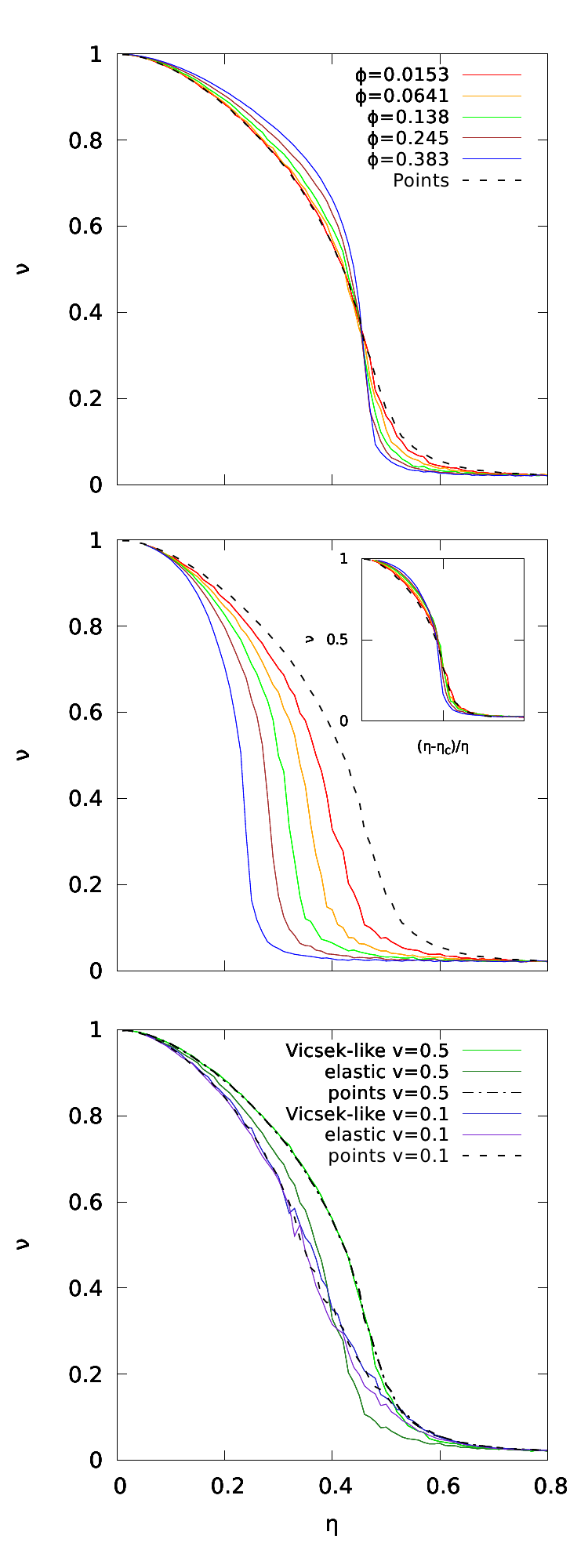}
}
\caption{Order parameter as a function of the noise. Top: Active disks with Vicsek-like bouncing rule. Middle: Active disks with pseudo-elastic bouncing rule, inset: order parameter as a function of the  noise rescaled with respect to critical noise ($\eta_c$). In all cases $N=2000$, $L=32$ and $v=0.5$. The same color code is applied in both top and middle panels.  The dashed line represent results for the point-like Vicsek model. Bottom: $\nu$ versus $\eta$  for $v=0.1$ (blue/violet lines) and $v=0.5$ (green lines) for point-like Vicsek model (dashed) and active disks at $\phi=0.0153$ with both bouncing rules.}
\label{packing}     
\end{figure}

 In order to follow the order-to-disorder transition, we compute the order parameter $\nu$ 
 (as in eq.\ref{ordpar})  as a function of the noise strength. 
 The top panel  in Fig. \ref{packing} represents $\nu$ versus $\eta$ 
 for disk-like Vicsek particles interacting via Vicsek-like bouncing rule (coloured continuous lines) as compared to 
  point-like Vicsek particles (dashed black line).
 As expected, not only the 
 the location of the order-disorder transition is closer to that of point-like Vicsek particles as $\phi$ approaches to 0
 but also the transition becomes sharper as  the packing fraction increases.

Similarly, when hard-disk Vicsek particles interact via  the pseudo-elastic bouncing rule (middle panel) 
not only the 
 the location of the order-disorder transition is closer to that of point-like Vicsek particles as $\phi$ approaches to 0
 but also the transition becomes sharper as  the packing fraction increases.
  As shown in the  inset of figure \ref{packing},  
 scaling the noise strength with the critical noise, $\eta_c$, all curves collapse onto a master curve 
overlapping with the Vicsek point-like   results.

Therefore, we conclude that neither excluded volume interactions nor the collision rules are enough to change
 the location and the nature of the order-disorder transition in the Vicsek model.
 While the pseudo-elastic bouncing rule   shifts the transition for increasing $\phi$ towards smaller noise strengths (middle panel),  
when particles interact Vicsek-like (top panel)
  the excluded volume interactions only affect the sharpness of the transition.

The results presented so far have been obtained with a particles' speed of $v=0.5$.
In order to establish the role played by the particles' velocity 
on the order-disorder transition for the two collision rules, we study a system of particles 
whose propulsion speed is $v=0.1$.
  In the bottom panel of Fig. 4 we compare the results obtained at v = 0.5 and v = 0.1 for the Vicsek-like and pseudo elastic bouncing rules for a system of
self-propelled disks.

At high velocities (v = 0.5), the order-disorder transition is affected by the bouncing rule, with the VIcsek-like rule in better agreement with the point-like Vicsek model. On the contrary,
for lower values of $v$, this effect is much weaker and the results obtained with both bouncing rules 
are indistinguishable with respect to those for the point-like 
Vicsek particles. 
Therefore, to underline the effect of the applied bouncing rule, from now onward we will 
only consider particles propelling at $v=0.5$.




\subsection{Point-like Vicsek particles with fixed obstacles}
\subsubsection{Aligning bouncing rule}\label{res2}

\begin{figure}[h!]
\centering
\resizebox{0.35\textwidth}{!}{%
\includegraphics{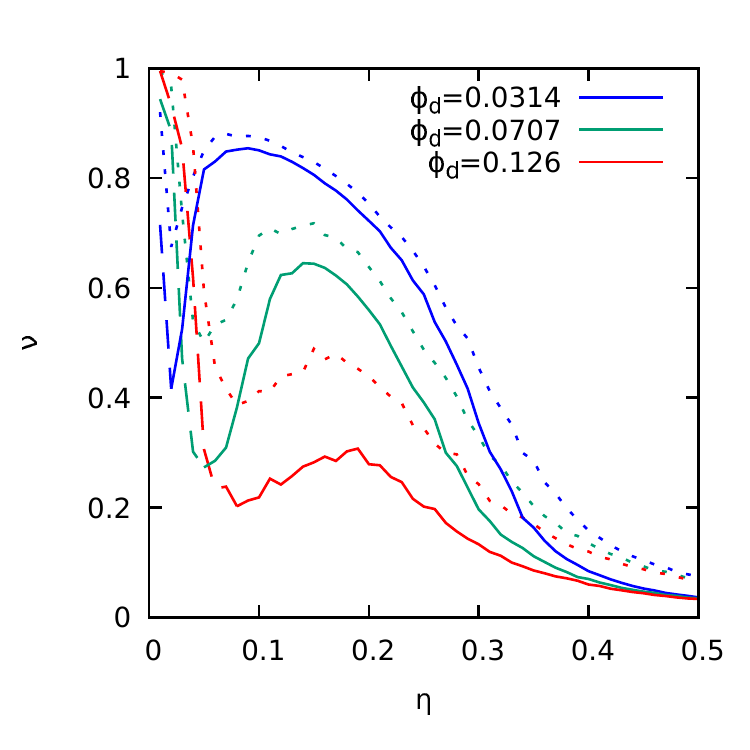}
}
\caption{Order parameter as a function of the noise strength for various obstacles packing fractions.  
Continuous lines: $N=8000$, $N_{d}=256$, $L=64$, $v=0.5$. Dotted lines: the same but with $L=32$. 
Dashed lines (at low $\eta$) correspond to finite-size effects.
}
\label{finsize}     
\end{figure}

We carry out simulations of a system composed by point-like Vicsek particles moving at v = 0.5 
embedded in a random array of fixed obstacles at different obstacle packing fractions, $\phi_d$. 
The obstacles packing fraction is changed by increasing their diameter, $\sigma_d$, while keeping constant their number, N$_d$. 
The results of the order parameter, $\nu$, as a function of the noise strength, $\eta$, are presented in figure \ref{finsize}.     
  


For large values of $\eta$  the system is disordered, as shown by the small value of the  order parameter: 
if particles form a cluster, each cluster presents a different direction of motion (while 
particles will be constantly exchanged between clusters, as shown in the right panel of Fig. 6). 
Decreasing the noise strength  particles tend to align forming larger clusters that swim in the same direction,  
reason why the order parameter $\nu$ starts  increasing. 

 \begin{figure}[h!]
\centering
\resizebox{0.21\textwidth}{!}{%
\includegraphics{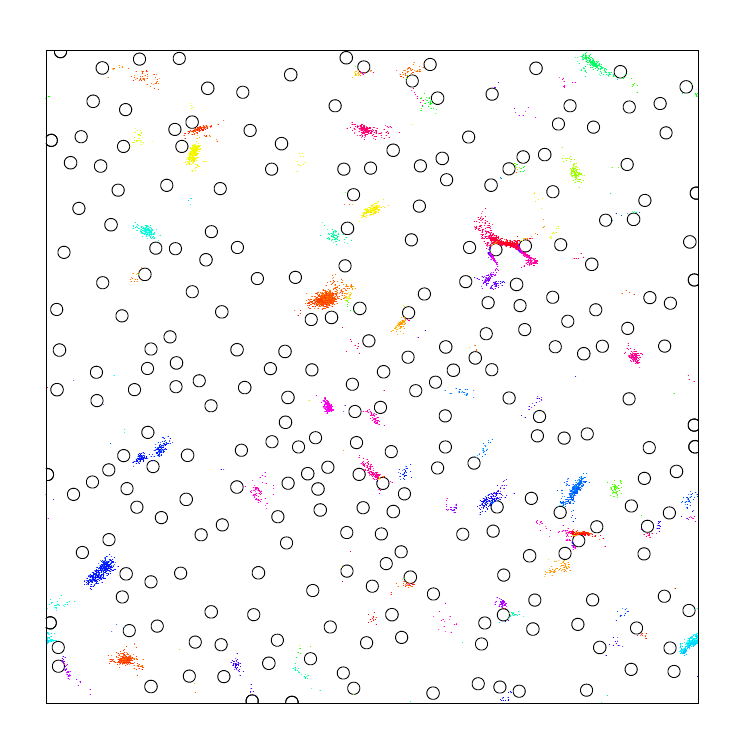}
}
\resizebox{0.245\textwidth}{!}{%
\includegraphics{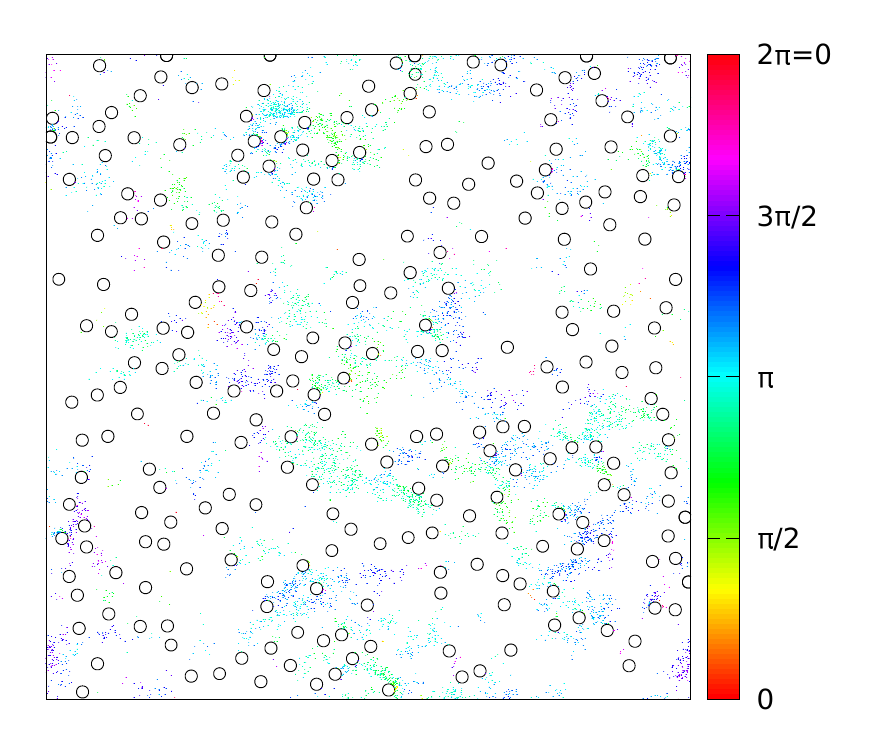}
}
\caption{Two snapshots from the green curve of figure \ref{finsize} ($\phi_d=0.07$). The color code  represents each particle's direction. Left, $\eta = 0.05$:  each cluster has an independent direction of motion (inducing a small  order parameter).  Right, $\eta = 0.14$:  clusters interact with each other thus increasing order.}
\label{diagrui}     
\end{figure}

However, unlike in the absence of obstacles, the order parameter reaches a maximum value for non-zero noise strength. 
As  already reported by Chepizhko and Peruani \cite{chepizhko2015active}, 
obstacles favour the formation of  clusters   as compared to what happens in the bulk system. 

When decreasing the noise strength even further, $\nu$ starts decreasing: 
This is due to the fact that,  
below a certain noise level, clusters do not interact any more and 
their collision with  obstacles only randomise their swimming directions
(as shown in the left panel of Fig. 6).
At the lowest noise levels (dashed lines in Fig. 5) further decreasing of $\eta$ leads to 
 an increase of $\nu$, which corresponds to the formation of a single cluster in the system: in this regime, 
 the numerical results are affected by the system's finite size. 


This behaviour is independent  on  the values of 
the obstacle packing fractions $\phi_d$ (figure \ref{finsize}).
Interestingly, the lower  the $\phi_d$ 1) the larger the value of $\nu$ (higher order) in the system at every noise level; 
2) the more pronounced the shift to the right of the maximum of $\nu$
(in the limit of no obstacles,  we should recover the results obtained for the bulk Vicsek). 

\begin{figure}[h!]
\centering
\resizebox{0.35\textwidth}{!}{%
\includegraphics{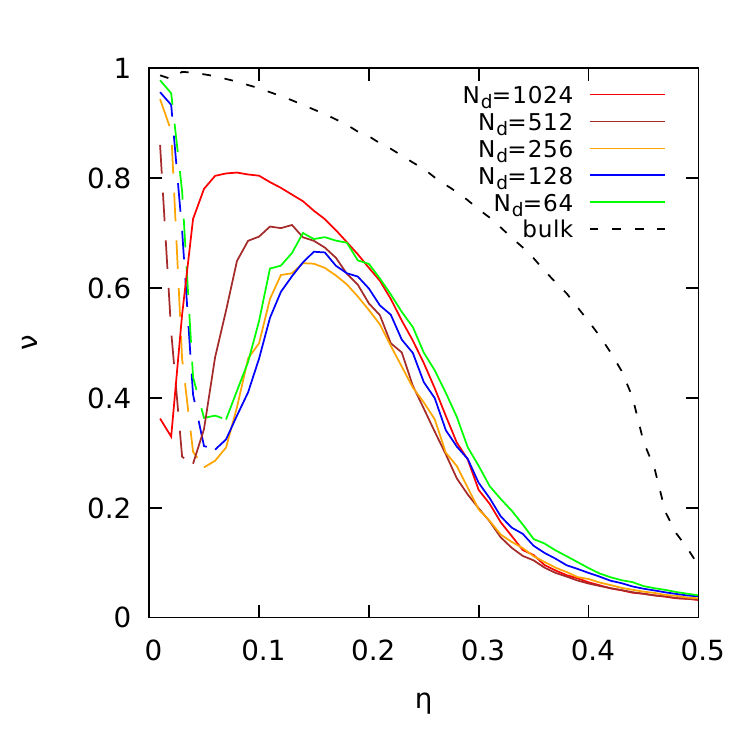}
}

\caption{ Order parameter as a function of noise, for different  obstacles size at the same packing fraction. $N=8000$, $L=64$, $v=0.5$, $\phi_d=0.0707$. The dashed black line are the results for the bulk Vicsek model, whereas the dashed coloured lines (at low $\eta$) correspond to finite-size effects.}
\label{nd}     
\end{figure}

In order to  understand whether the results might be affected by the obstacles' size, we keep 
their packing fraction fixed at $\phi_d=0.0707$ while systematically changing their diameter 
( thus the  obstacles' number $N_d$)  
and compare to the bulk Vicsek equivalent (at the same $L=64$),  
as shown in figure \ref{nd}.


While the location of the maximum   of $\nu$  
at lower values of $\eta$ is clearly affected by the the obstacles' size, 
as soon as $\eta \ge 0.2$  $\nu$  does not depend on $N_d$ (while it depends on $\phi_d$, as shown in fig. \ref{finsize}).
When $64 \le N_d \le 256$ ($ 2.4 R \ge \sigma_d \ge 1.2 R$)  all curves overlap for every noise level:
the  effect of  obstacles on clustering is the same independently on their diameter, always larger than the interaction range $R$. 

  As soon as  $\sigma_d$ becomes smaller than $R$ ($N_d > 256$),   the maximum    
   of $\nu$ starts shifting towards lower values of $\eta$.  
One can extrapolate that in the limit of  $\sigma_d \to 0$ the maximum  shifts to $\eta \to 0$.





\begin{figure}[h!]
\centering
\resizebox{0.4\textwidth}{!}{%
\includegraphics{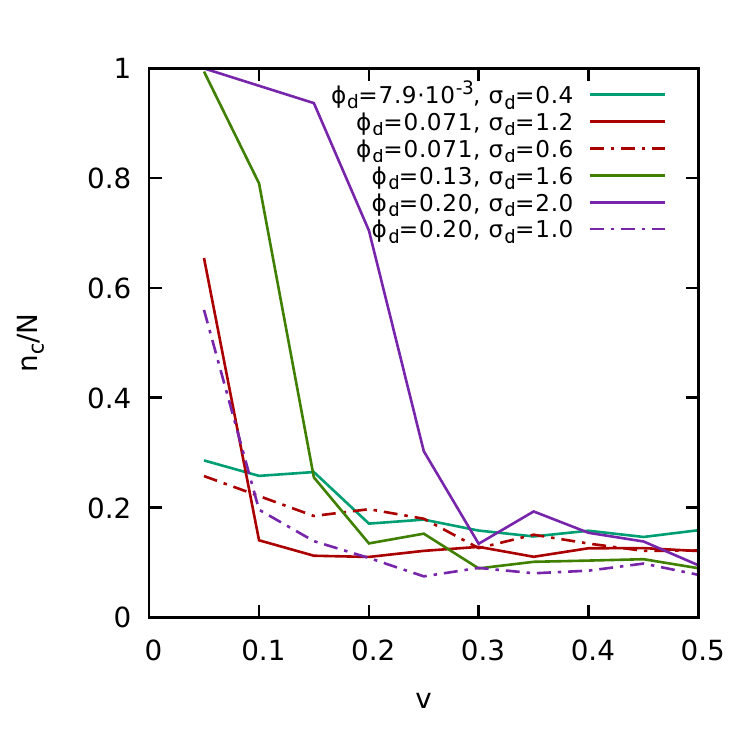}
}

\caption{Average number of particles in the largest cluster ($n_c$) normalised by the total number of Vicsek particles ($N$) as a function of the speed $v$.  $N=8000$, $\eta=0.1$,  $L=64$. We combine different packing fractions of obstacles and different obstacle diameters. We see that in some cases the largest cluster grows when decreasing $v$.}
\label{cl}     
\end{figure}

So far we have been simulating a system with particles propelling at the constant speed of $v=0.5$.
In order to assess the relevance of particles' speed $v$ on clustering 
in the presence of random fixed obstacles at different packing fractions, 
we study systems
 with   $\phi_d \in [7.9 \times 10^{-3},0.2]$ and $\sigma_d\in [0.4,2.0]$
 where particles propel at several speeds  while keeping    $\eta=0.1$ fixed. 
 As shown  in   figure \ref{cl}, we 
  compute the average number of particles in the largest cluster $n_c$ (in steady state) as a function of $v$, 
that  is a relevant quantity to study the collective behaviour,  corresponding to the maximum size  reached by a cluster  
 before breaking in a collision with an obstacle. 

  

\begin{figure}[h!]
\centering
1)\resizebox{0.205\textwidth}{!}{%
\includegraphics{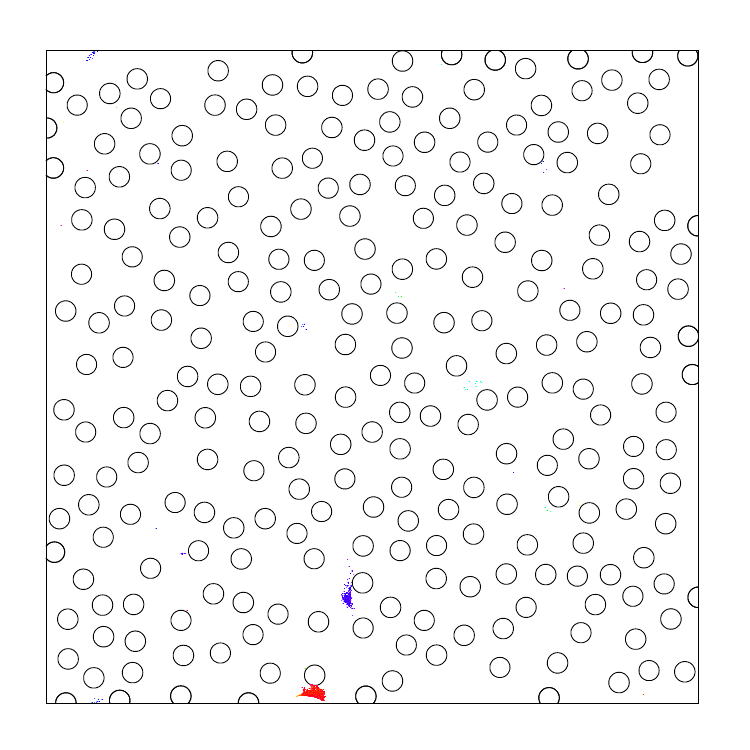}
}
2)\resizebox{0.24\textwidth}{!}{%
\includegraphics{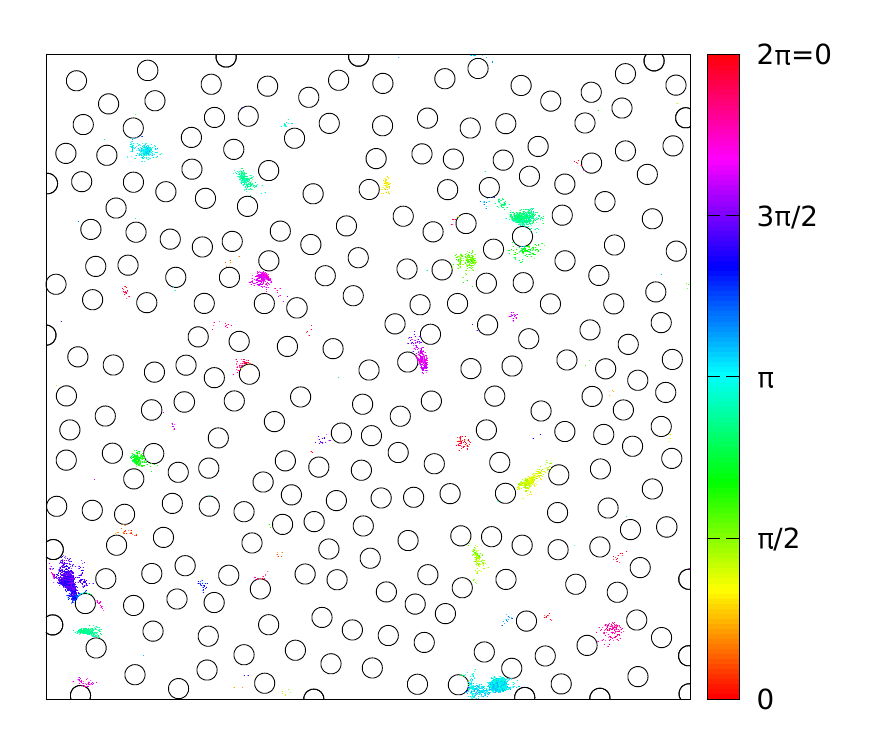}
}
3)\resizebox{0.205\textwidth}{!}{%
\includegraphics{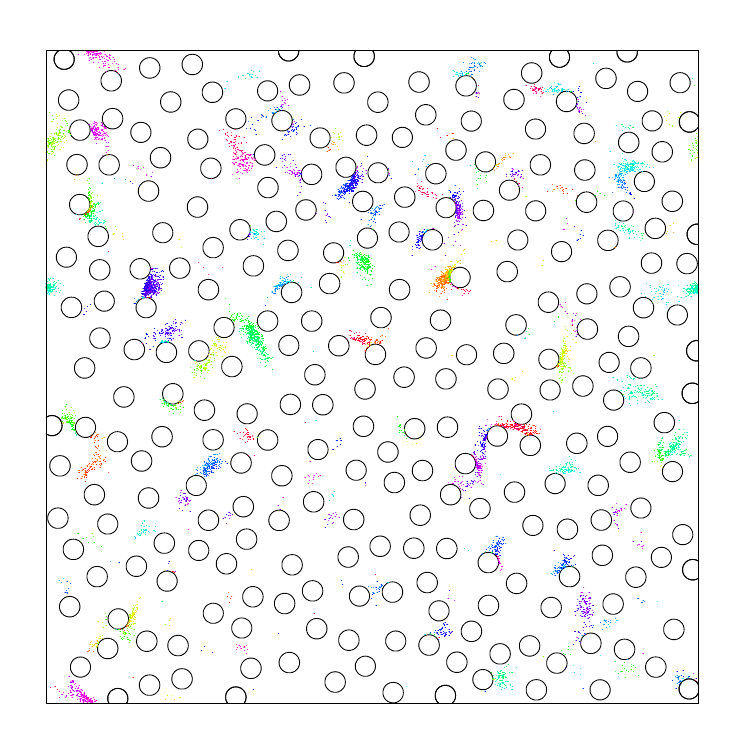}
}
4)\resizebox{0.24\textwidth}{!}{%
\includegraphics{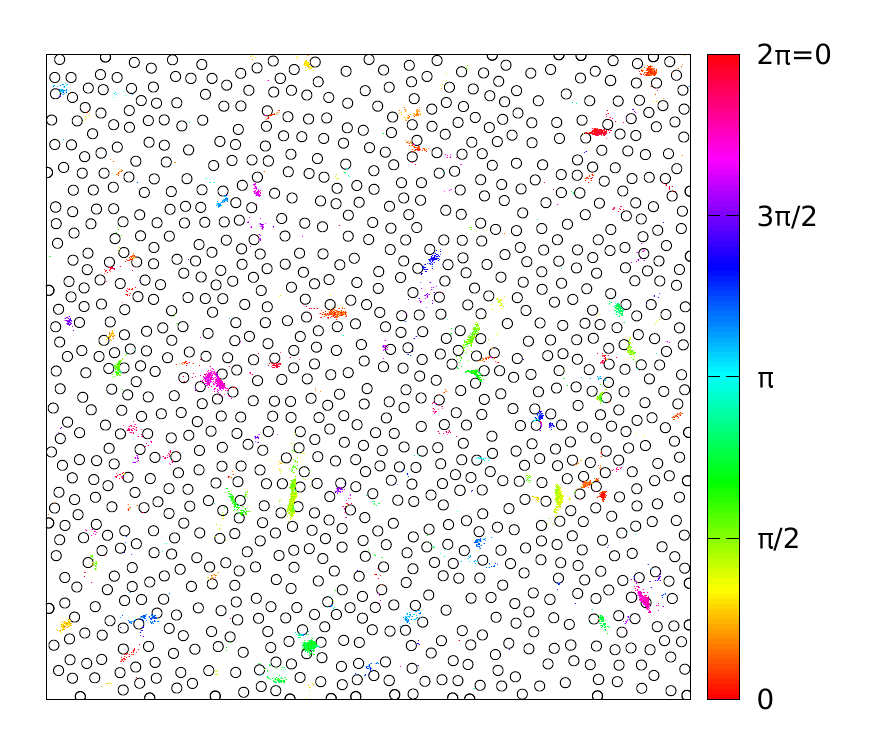}
}

\caption{Snapshots taken in the steady state when $\sigma_d=2$, $\phi_d=0.20$ (figure \ref{cl}) and $v=0.15$ (1), $v=0.25$ (2) and $v=0.5$ (3), respectively. (4) represents the case when  $\sigma_d=1$, $\phi_d=0.20$ and $v=0.15$ (same $\phi_d$ and $v$ as (1 ). In the first three ones, we see that clusters are denser and have more particles as we decrease speed ($v$), while snapshot 4) shows that this effect is suppressed as obstacle size is lower, even for the same obstacles packing fraction.}
\label{diag}     
\end{figure}

We observe that 
 whenever the speed $v$ is high, independently on $\phi_d$, $n_c/N$ is low given that 
    clusters have good probabilities of breaking  when encountering an obstacle: 
 as shown in the  panel 3 of figure \ref{diag}), 
  many small clusters are formed. 

However, when the speed starts decreasing, the behaviour is strongly affected both by   $\phi_d$ and $\sigma_d$:
 $n_c/N$ increases especially for the largest    values of $\phi_d$. 
This effect is more pronounced when $\sigma_d$ is  larger than $R$ (continuous lines in figure \ref{cl}).  
This is due to the fact that 
   particles tend to be concentrated in a small number of clusters,  each   containing many particles; 
  cluster breaking is more difficult    (panels 1 and 2 of figure \ref{diag}) given that even when  particles are 
  scattered due to the collision with an obstacle they can still interact after the collision, especially if their distance is  smaller than $R$.

    At the lowest value of $\phi_d$ and $\sigma_d$ (continuous light-green curve in figure \ref{cl}), small clusters are formed in the system. 
  A similar behaviour is observed for every $\phi_d$ whenever  $\sigma_d$ is smaller than $R$ (dotted-dashed lines in figure \ref{cl}), 
  being  the effect of $v$ on the 
  cluster size not detectable.

 The cluster size is also affected by tuning the obstacles' size.  
  In panel 4 of figure \ref{diag}  the 
 diameter of the obstacle is set to  half the size of the obstacles presented in panel 1 of the same figure, 
 with particles having  the same  speed. 
 While in the  former case the system is characterised by many clusters  (a low value of $\nu$),  in the latter 
the system consists of  few large clusters.

 To conclude, the cluster size can be tuned by either tuning the Vicsek's particles velocity  or the diameter of the obstacles 
 (depending on the ratio between   $\sigma_d$ and $R$). 

\subsubsection{Elastic bouncing rule}\label{res3}


To highlight the role played by the bouncing rule, we 
have studied a system of $N=2000$ Vicsek point-like particles with $L=32$ and $v=0.5$
 interacting with   $\phi_d=0.0707$ ($N_d=64$) obstacles via 
pseudo-elastic bouncing rules ("el" brown dashed line in figure 10) and compared it to the same system 
interacting via an aligning rule ("Vi" dark green dashed line in figure 10, replotted from figure 5 (dotted green line)). 
\begin{figure}[h!]
\centering
\resizebox{0.4\textwidth}{!}{%
\includegraphics{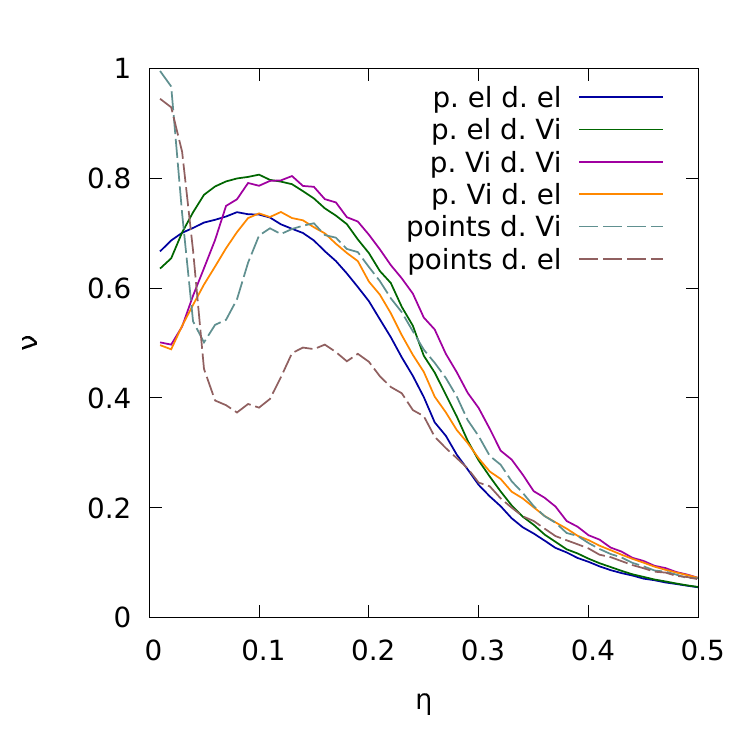}
}

\caption{ Order parameter versus noise for   $N=2000$ Vicsek point-like particles with $L=32$ and $v=0.5$ interacting with obstacles at 
 $\phi_d=0.0707$ ($N_d=64$). 
Dashed lines are for point-like Vicsek particles interacting with obstacles 
via  aligning (Vicsek-like) (Vi) or elastic (el) rules. 
Continuous lines correspond to a system of $\phi=0.0153$   finite-size Vicsek disks: `p.' denotes the bouncing rule between particles and `d.' the bouncing rule with obstacles.}
\label{hdbound}     
\end{figure}

Comparing the dashed lines in figure 10, we conclude that the  overall collective features of $\nu$ versus noise
are not altered by the chosen bouncing rule with obstacles, whether   aligning  ("Vi" dark green dashed line) or  elastic ("el" brown dashed line). 
In either case, the system is disordered (low value of $\nu$) for high noise; becomes more ordered when decreasing $\eta$ 
up to the point of reaching a maximum value of $\nu$   approximately at the same $\eta$, given that  
obstacles favour the formation of  clusters. 
Interestingly, an elastic bouncing rule is responsible for the formation of less ordered aggregates (lower $\eta$ maximum)
 When decreasing the noise strength even further, $\nu$ starts decreasing,  
since  clusters do not interact any more and 
their collision with  obstacles only randomise their swimming directions
At the lowest noise levels (dashed lines in Fig. 10) further decreasing of $\eta$ leads to 
 an increase of $\nu$, which corresponds to the formation of a single cluster in the system: in this regime, 
 the numerical results are affected by the system's finite size.

Therefore, we conclude that even though  
 the aliging bouncing rule enhances ordering among the particles, the  
 maximum of the order parameter is an  intrinsic feature of the presence of obstacles, 
 being present when particles collide with either an aligning or an elastic bouncing rule with the obstacles.


\subsection{Disk-like  Vicsek particles with fixed obstacles: aligning versus pseudo-elastic bouncing rule}\label{res4}


Having defined in section \ref{Sec:Simulation_Details}
 two bouncing rules between active disks and other two bouncing rules for collisions between an active disk and an obstacle, 
  one might ask how 
the active disk-obstacle bouncing rule affects the collective behaviour of the active disks, and under which conditions the inter-particle bouncing rule is more relevant.
We prepare  a system of $N=2000$ disks propelling at a speed of $v=0.5$ (the same as for point-like particles in figure 10) 
 with a packing fraction $\phi=0.0153$ in a simulation box of $L=32$. In the same box we add $N_d=64$ obstacles placed 
 randomly with a packing fraction $\phi_D=0.0707$ ($\sigma=0.1$ for  disks
 and $\sigma_d=1.2$ for  obstacles).
 The reason for choosing active disks  one order of magnitude smaller than obstacles, is dictated by the fact that we expect them to show 
a similar behaviour of the order parameter with respect to  Vicsek point-like particles interacting with the same obstacles.
We have studied four different cases presented in sections \ref{s2} and \ref{s3}, combining the different collision rules: 
1) elastic bouncing rule between particles and elastic bouncing rule between a particle and an obstacle ("p. el d. el", blue continuous line in figure 10)
2) elastic bouncing rule between particles and aligning bouncing rule between a particle and an obstacle ("p. el d. Vi", dark green continuous line in figure 10)
3) aligning bouncing rule between particles and aligning bouncing rule between a particle and an obstacle ("p. Vi d. Vi", magenta continuous line  in figure 10)
4) aligning bouncing rule between particles and elastic bouncing rule between a particle and an obstacle ("p. Vi d. el", orange continuous line in figure 10)




 As shown for all continuous lines of  figure \ref{hdbound}, 
the behaviour of the system consisting of finite size disks  is similar to the one consisting of point-like particles, independently 
on the chosen bouncing rules.
The only difference between finite-size and point-like particles can be detected for very low values of  $\eta$, 
given that  the increase of $\nu$    is absent for active disks. 
This is due to the fact that  excluded volume  prevents to find only one cluster of active disks in the simulation box.

 
As for  point-like particles, the system of active disks is more ordered when  disks interact via aligning bouncing rules with the obstacles, 
independently on the chosen inter-particle bouncing rule 
 (as shown by comparing the blue (elastic) and orange (Vicsek)  lines
 to the dark green dashed line (point-like particles)  for the particle-obstacle  elastic bouncing rule 
 and the green (elastic) and magenta (Vicsek) lines  
 to the brown dashed   line (point-like particles) for the  particle-obstacle  aligning bouncing rule).
However, in all cases, excluded volume prevents  clusters formation, thus enhancing global order.

 
 Therefore, while near the maximun of $\nu$ the disk-obstacle bouncing rule is  crucial 
 for the system's features,  at the lower/higher values of $\eta$ the disk-disk
  bouncing rule is the most important one. 
  In the former case, 
   given that when global polar order reaches a maximum value
  all particles follow a similar direction, particle-particle collisions are reduced, reason why  
   collisions with the obstacles are more relevant.
 In the latter case, when the noise strength is high   
  particles do not form a cluster and frequently collide among them, whereas
  when the noise strength is low 
 particles form clusters that move in different directions and the frequent 
   collisions between two clusters is the reason  why particle-particle bouncing rules dominate. 

\begin{figure}[h!]
\centering
\resizebox{0.4\textwidth}{!}{%
\includegraphics{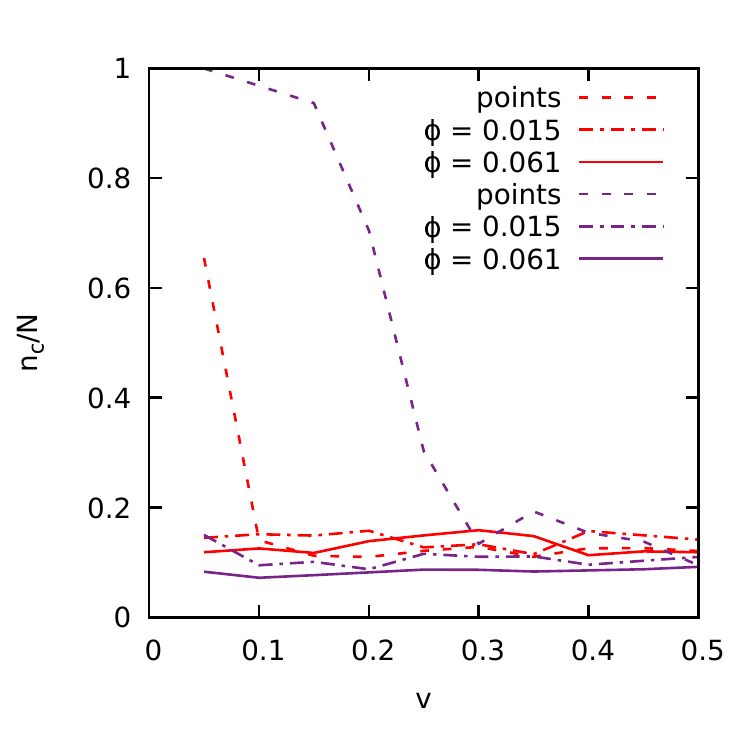}
}

\caption{Average number of particles in the largest cluster ($n_c$) normalised to the total number of Vicsek particles as a function of the speed $v$.  In red, $\phi_d= 7.1 $ $10^{-2}$ 
whereas in purple $\phi_d=0.20$. $N=8000$ active particles,  $\eta=0.1$ and $N_d=256$ obstacles in a simulation box $L=64$ (in either $\phi_d$ we have considered $\sigma_d$ larger than $R$). 
 In both cases, the dashed lines  correspond to Vicsek points, the dotted-dashed to $\phi=0.015$ and the continuous lines to $\phi=0.061$. 
Active disks follow a Vicsek-like bouncing rule between them and with the obstacles. }
\label{cl2}     
\end{figure}

As shown  in   figure \ref{cl} for point-like particles, we now  
  compute the average number of particles in the largest cluster $n_c$ (in steady state) as a function of $v$ (figure \ref{cl2}), 
corresponding to the maximum size  reached by a cluster  
 before breaking in a collision with an obstacle. 
 
We study systems of active disks  with a particles' speed ranging from 
0.1 up to 0.5, setting   the same noise strength $\eta=0.1$, whose bouncing rules are aliging 
either among them or with obstacles. 
The systems are prepared at xs different  packing fractions of $N_d=256$  obstacle/$N=8000$ active particles in a simulation box of $L=64$.
We choose to examine a lower $\phi=0.015$ (dotted-dashed lines) and a higher $\phi=0.061$ (continuous lines) 
packing fraction,  comparing them both to the point-like case (dashed lines).
Given that the   packing fraction is computed  by changing $\sigma$ and $\sigma_d$ while keeping  $N$ and $N_d$ fixed, 
we compare the case with   $\phi_d= 7.1 $ $10^{-2}$ (all red curves in figure \ref{cl2}) 
 to the one with  $\phi_d=0.20$ (all purple curves in figure \ref{cl2}).

Differently from the point-like system, the disks' largest cluster size  does not change with the self-propulsion speed with this set of bouncing rules 
and for these four combinations of packing fractions. The emergence of a big cluster is suppressed by the particles' concentration  limit
 imposed by the   hard disks'  excluded volume.





\subsection{Disk-like  Vicsek particles with moving obstacles: aligning versus pseudo-elastic bouncing rule}\label{res5}

In a recent study Llanes and coworkers\cite{yllanes2017many} have demonstrated that many non-aligning passive disks 
are more effective at 
destroying the ordered phase in a suspension of active Brownian particles with an aligning interaction.
Our goal is to study the effect played by passive moving particles  
in a suspension of self-propelled Vicsek disks, when changing the drag coefficient of the passive particles. 
 
\begin{figure}[h!]
\centering
\resizebox{0.4\textwidth}{!}{%
\includegraphics{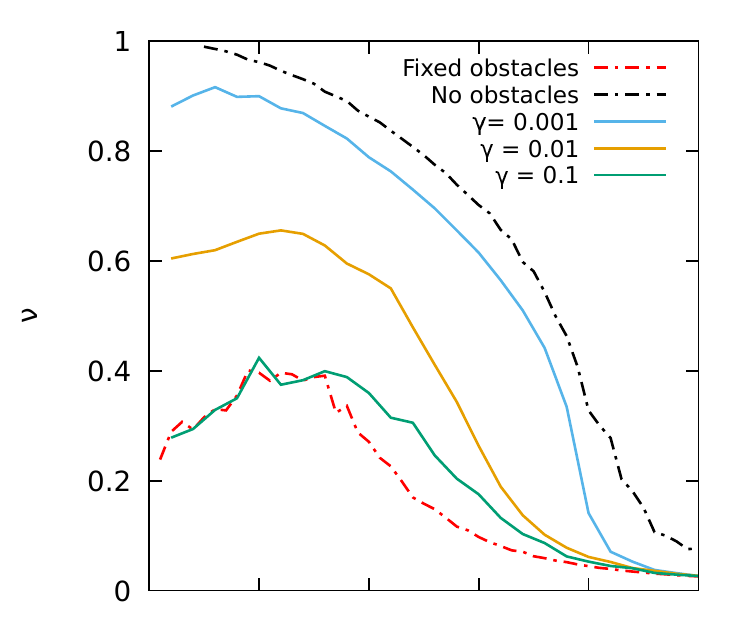}
}
\resizebox{0.4\textwidth}{!}{%
\includegraphics{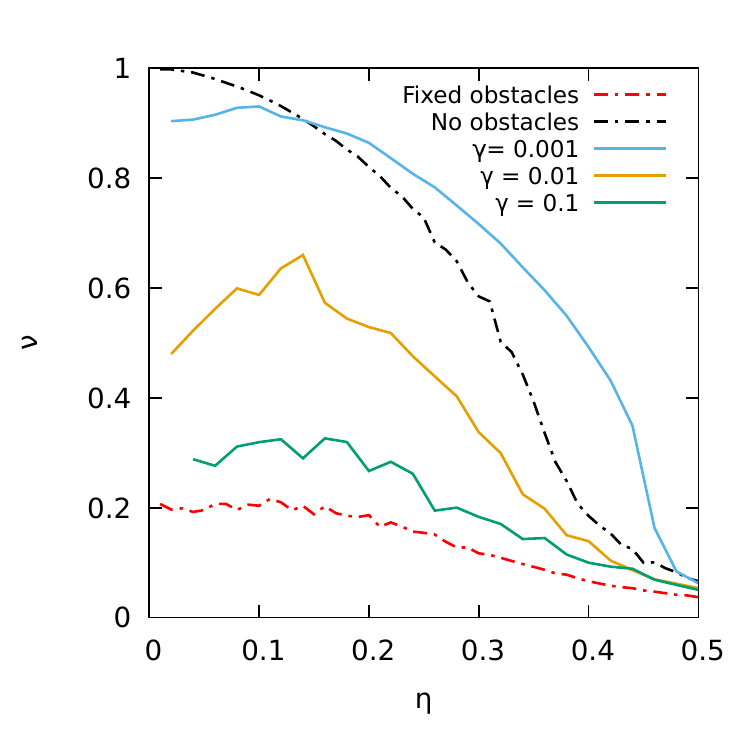}
}

\caption{  Order parameter versus noise for $N = 8000$, $L = 64$, $N_d = 256$, $\phi_d =0.0707$, $\phi = 0.0153$: top, $v=0.5$ and $\Delta t'=0.1$. Bottom, $v=0.1$ and $\Delta t'=0.5$. }
\label{diss}     
\end{figure}
We  now consider a dilute suspension of $N=8000$ Vicsek disk-like particles 
at $\phi=0.0153$ ($L=64$) in the presence of $N_d=256$ moving obstacles ($\phi_d=0.0707$) with  $\sigma /\sigma_d=1/10$ 
\footnote{To establish the  effect of  $\sigma /\sigma_d$, we have  performed 
simulations with  $\sigma =\sigma_d$  ($N=N_d$), where  
 the effect of the passive
 particles in destroying order was  quite small even for a large value of the drag 
(data not shown).}
 Passive particles  motion is mainly generated by the inertia due to the collisions with  active particles, being 
  modelled   via an over-dumped dynamics, with a drag coefficient $\gamma$ 
that restores the particle to rest. 
At intervals of $\Delta t'$ (smaller than $\Delta t=1$), each moving obstacle 
is slowed down according to eq.3 in   section \ref{s4}
The passive disk-passive disk bouncing rule is set to be 
  elastic (rather than  aligning); whereas 
 the passive disk- active disk  and active disk- active disk one 
is a pseudo-elastic one.

 We compute the order parameter $\nu$ at different noise strength with $\gamma$ 
ranging from 0.001  (continuous blue lines in fig. \ref{diss}), 0.01 (orange lines) to 0.1 (green lines), 
comparing to the system with the same packing fraction of fixed obstacles (dashed-dotted red lines) 
and to the bulk system with no obstacles (dashed-dotted black lines).
To study the role played by the particles' velocities, we consider 
  disks propelling at $v=0.5$ (thus  $\Delta t'=0.1$) (top panel of figure  \ref{diss}),  and 
  propelling at $v=0.1$ (thus  $\Delta t'=0.5$) (bottom panel of figure  \ref{diss}).

 To start with we observe
that, independently on the propulsion speed,  the system is more disordered when 
moving obstacles are strongly over-damped ($\gamma=$0.1, green lines in fig.\ref{diss}) 
approaching results obtained  for fixed obstacles (dashed - dotted red lines in fig.\ref{diss}). 
\begin{figure}[h!]
\centering
\resizebox{0.21\textwidth}{!}{%
\includegraphics{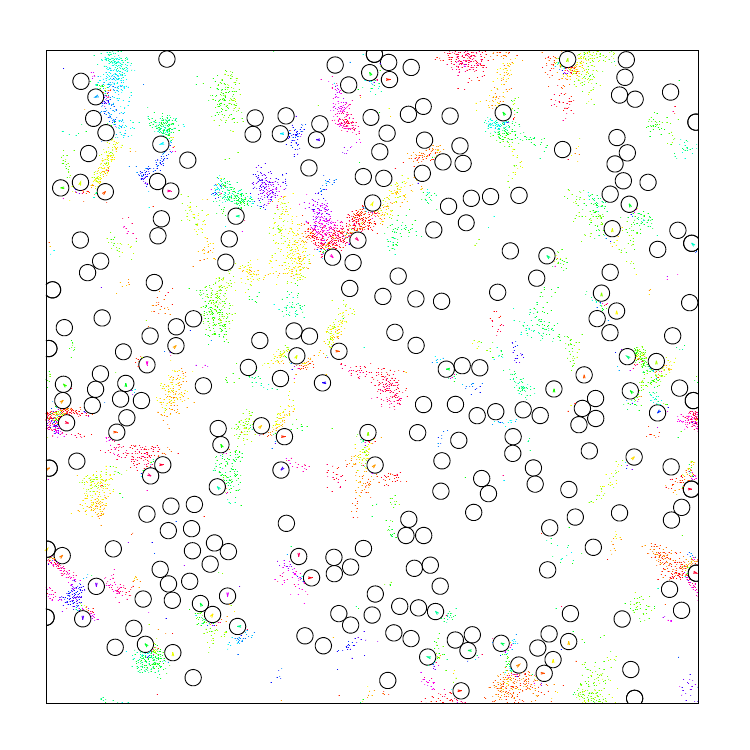}
}
\resizebox{0.245\textwidth}{!}{%
\includegraphics{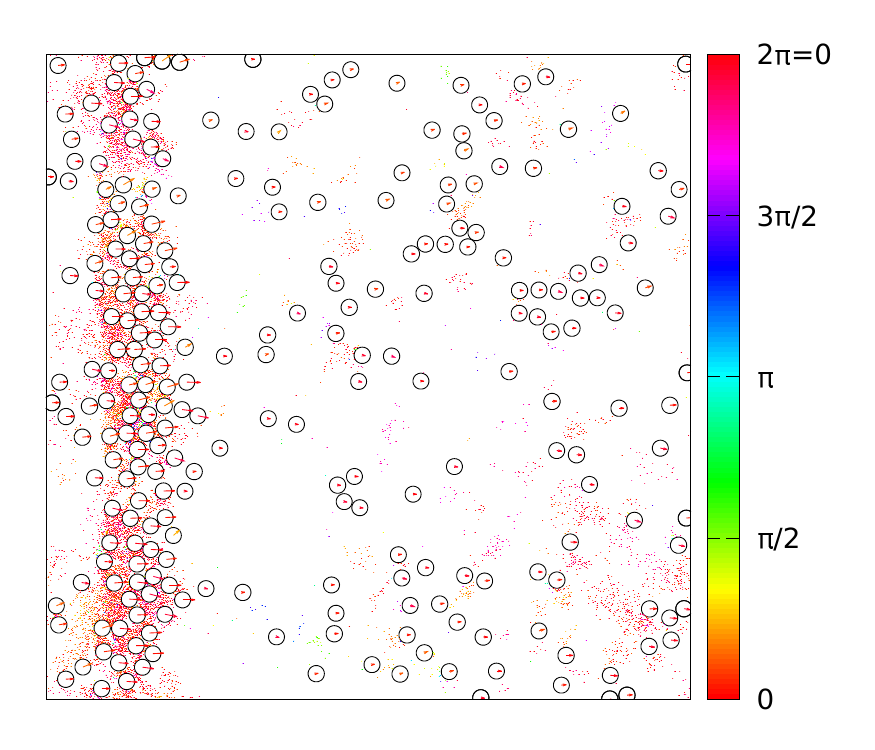}
}
\caption{Snapshots representing: left)  high drag ($\gamma=0.1$) and right) low drag ($\gamma=0.001$)
 $N=8000$, $\phi=0.0153$, $L=64$, $v=0.5$, $\eta=0.1$, $N_d=256$, $\phi_d=0.0707$ and $\Delta t'=0.1$. When passive particles' drag is high, we see cluster formation of active particles (left), like in the fixed obstacles case. When drag is low, we see that passive particles are carried away by active particles in a band (right).
} 
\label{diagdiss}     
\end{figure}

When studying the system with Vicsek particles propelling at  $v=0.5$ (top panel in fig.\ref{diss}) we observe
that when $\gamma$ is high we recover results obtained in the fixed obstacles case: active clusters form clusters 
(left-snapshot in figure   \ref{diagdiss}). However,  when
 decreasing $\gamma$, 
ordering increases up to the point of almost recovering the results when no-obstacles are present (dashed-dotted black lines in fig.\ref{diss}).
At the lowest  $\gamma$ (right panel in  figure   \ref{diagdiss})
 active particles tend to move in an order band. This leads 
to a peculiar effect of passive obstacles being dragged away by the active disks. 
In these cases,  passive mobile obstacles
 are not capable of destroying the order within the active particles.

When the Vicsek's particles speed is lower (bottom panel in fig.\ref{diss}), 
the system stays quite ordered even at lowest noise 
 (as compared to the fixed obstacles case), 
even though $\nu$ is always lower than the one evaluated for particles 
propelling at higher speed (top panel in the same figure).  

\begin{figure}[h!]
\centering
\resizebox{0.21\textwidth}{!}{%
\includegraphics{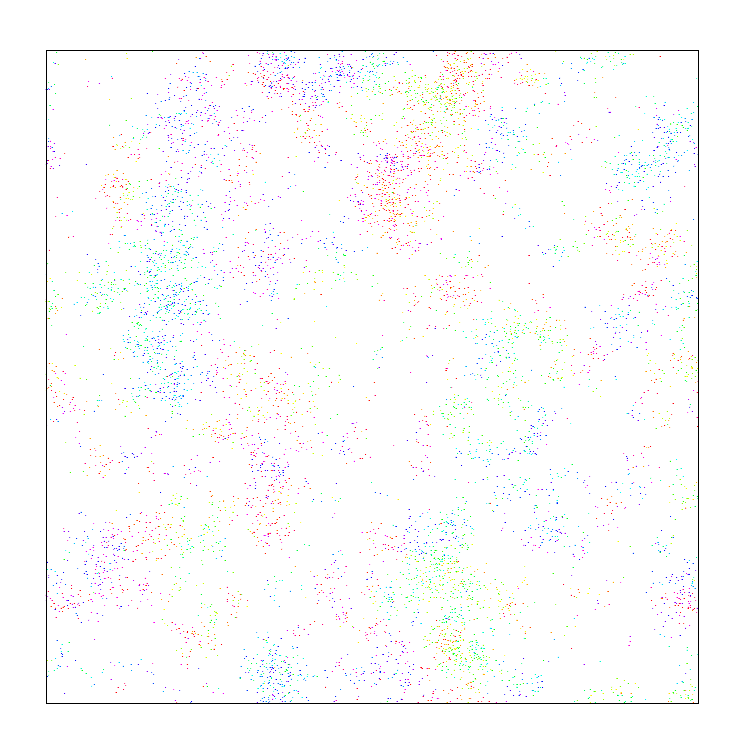}
}
\resizebox{0.245\textwidth}{!}{%
\includegraphics{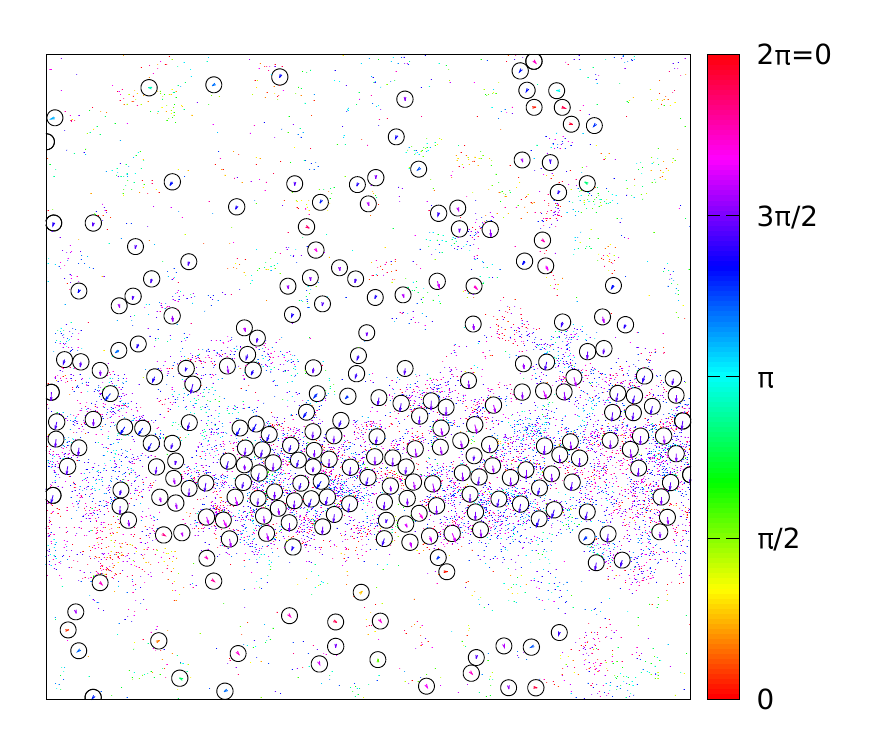}
}
\caption{Snapshots representing: left) bulk active disks; right) active disks with $\phi_d=0.0707$ ($N_d=256$) moving obstacles ($\gamma=0.001$ and $\Delta t'=0.5$). 
Parameters are $N=8000$, $\phi=0.0153$ and $L=64$, $v=0.1$, $\eta=0.4$. In the right snapshot we see that in the passive low drag passive particles case active particles are more aggregated than in the bulk case (left), and this produces the counterintuitive enhance of order for some values of $\eta$.
}
\label{intnoise}     
\end{figure}

Counterintuitively, we observe   that  
 when $\gamma\to 0$,  active particles in the presence of mobile obstacles (continuous blue line) 
 are more ordered than when no obstacles are present (dashed-dotted black line). 
Figure \ref{intnoise} represents a direct comparison between a snapshot of the bulk system (left panel) 
and one corresponding to the light blue curve in the bottom panel  of figure\ref{diss} (right panel), both taken for 
  $\eta=0.4$. 

At the lowest  $\gamma$ and for  low and intermediate noise levels
 active particles tend to move in an order band. This leads 
to a peculiar effect of passive obstacles being dragged away by the active disks. 
However, differently from the case of higher propulsion speed, 
active
particles (propelling at a low speed) are trapped by the presence of the obstacles, thus forced to be close to one another. 
Clearly,    obstacles   play the role of aggregating active particles, thus enhancing global order with respect to 
the bulk case. 

Therefore, depending on the active particles' speed, passive particles might (when the speed is low) 
or might not (when the speed is high) induce cluster formation of the active particles. 
A key role is also played by the ordered present in the bulk system.
 As already observed in figure 1-bottom panel, the system is more ordered when $v=0.5$, 
 at least for the chosen system size. 




\section{Conclusions} \label{Sec:Conclusions}

In our manuscript, we have presented a numerical method to  
simulate a two dimensional suspension of self-propelled  Vicsek disks evolving according to an event-driven dynamics.
To start with, we have studied the bulk system and  fully recovered 
the    order-disorder transition established for point-like Vicsek particles,   
independently on the chosen bouncing rule (aligning or pseudo-elastic). 
Therefore, neither excluded volume interactions nor the collision rules are enough to change
 the location and the nature of the order-disorder transition in the Vicsek model.
 While the pseudo-elastic bouncing rule   shifts the transition for increasing $\phi$ towards smaller noise strengths (middle panel),  
when particles interact Vicsek-like (top panel)
  the excluded volume interactions only affect the sharpness of the transition.

In order to establish the role played by the particles' velocity 
on the order-disorder transition for the two collision rules, we study a system of particles 
whose propulsion speed is lower.
While at high  velocities, the order-disorder transition is affected by the bouncing rule,
   with the VIcsek-like rule in better agreement with the point-like Vicsek model, when the particles' speed is lower
the results obtained with both bouncing rules 
are indistinguishable with respect to those for the point-like 
Vicsek particles.

Next, we use this methodology to simulate Vicsek particles in the presence of passive obstacles. First, we introduce point 
Vicsek particles in the presence of  fixed solid obstacles, considering aligning or elastic bouncing rules between the particles 
and the obstacles. 
 When obstacles are located in random positions, the  order-disorder
 transition is modified, showing  a maximum in the order parameter at non-zero noise strength. 
 As already  observed for a slightly different system \cite{chepizhko2015active}, we conclude that clusters formation 
  at low noise values  is the main reason of this counterintuitive result. 
 Analyzing the system of active disks with fixed obstacles, 
using two collision rules between disks and two bouncing rules between disk and obstacle, we observe that the disk-obstacle bouncing rule is dominant near the maximum of $\nu$, while the disk-disk one is more important for higher $\nu$ and in the $\nu\to 0$ limit.
  Even though   the aliging bouncing rule enhances ordering among the particles, the  
 maximum of the order parameter is an  intrinsic feature of the presence of obstacles, 
 being present when particles collide with either an aligning or an elastic bouncing rule with the obstacles.
 We also study the effect of the Vicsek  speed on cluster formation, 
concluding that the cluster size can be tuned by either tuning the Vicsek's particles velocity  or the diameter of the obstacles 
 (depending on the ratio between   $\sigma_d$ and $R$). 

Finally, we  introduce moving obstacles, whose  motion only depends on  collisions with  
active particles and their intrinsic drag.
We find  that  while 
highly damped  obstacles are the most effective in destroying the order between Vicsek disks, 
this order increases when approaching the no drag limit.  
Interestingly, for low self-propulsion speeds,  very low drag obstacles  induce  even more ordered states
 than what detected at the same concentration in a   bulk system.

\section*{Acknowledgements}

This work was funded by grants FIS2016-78847-P of the MINECO and the UCM/ Santander PR26/16-10B-2. C. Valeriani acknowledges financial support from a Ramon y Cajal Fellowship. R. Martinez acknowledges financial support from the FPI grant BES-2017-081108 by MINECO.

 We also thank Dr Jorge Ramirez for fruitful discussions.

%
%
\section*{Authors contributions}
R. Martinez was responsible of creating some of the models used, writing the code, running the programs and collecting the results. The rest of authors were involved in advising, as well as writing and revising the maniscript, where C. Valeriani and F. Alarcon have made the most important contributions. 
All the authors have read and approved the final manuscript.
%
 \bibliographystyle{ieeetr}
 \bibliography{referencias.bib}
%
%
%

\end{document}